\documentclass[journal=nalefd,manuscript=letter,layout=twocolumn]{achemso}

\usepackage{amsmath,amssymb,natbib,bm,graphicx,url,epsf,color}
\usepackage[english]{babel}
\usepackage{wasysym}
\usepackage{MnSymbol}
\usepackage{amsmath}
\usepackage{graphicx}
\usepackage{pdfpages}
\usepackage[colorinlistoftodos]{todonotes}
\usepackage[colorlinks=true, allcolors=blue]{hyperref}

\newcommand{\Vds}{V_{\mathrm{DS}}}
\newcommand{\VSi}{V_{\mathrm{Si}}}
\newcommand{\VgR}{V_{\mathrm{g1}}}
\newcommand{\VgT}{V_{\mathrm{g2}}}

\newcommand{\IR}{I_{\mathrm{R}}}
\newcommand{\IT}{I_{\mathrm{T}}}

\newcommand{\VR}{V_{\mathrm{R}}}
\newcommand{\VT}{V_{\mathrm{T}}}
\newcommand{\dIR}{\mathrm{d}I_{\mathrm{R}}}
\newcommand{\dIT}{\mathrm{d}I_{\mathrm{T}}}
\newcommand{\dIzero}{\mathrm{d}I_{\mathrm{0}}}
\newcommand{\dVR}{\mathrm{d}V_{\mathrm{R}}}
\newcommand{\dVT}{\mathrm{d}V_{\mathrm{T}}}

\makeatletter
\AtBeginDocument{\let\LS@rot\@undefined}
\makeatother

\makeatletter
\def\NAT@spacechar{}
\makeatother

\author{R. Delagrange}
\author{G. Le Breton}
\affiliation{Universit\'e Paris-Saclay, CEA, CNRS, SPEC, 91191 Gif-sur-Yvette cedex, France
}

\author{K. Watanabe}
\author{T. Taniguchi}
\affiliation{National Institute for Materials Science, Tsukuba, Japan
}
\author{P. Roulleau}
\author{P. Roche}
\author{F.D. Parmentier}
\email{francois.parmentier@cea.fr}
\affiliation{Universit\'e Paris-Saclay, CEA, CNRS, SPEC, 91191 Gif-sur-Yvette cedex, France
}

\title{Residual quantum coherent electron transport in doped graphene leads}

\begin{document}

\date{\today}

\begin{abstract}
\textbf{Recent low-temperature electron transport experiments in high-quality graphene rely on a technique of doped graphene leads, where the coupling between the graphene flake and its metallic contacts is increased by locally tuning graphene to high doping near the contacts. While this technique is widely used and has demonstrated its usefulness numerous times, little is known about the actual transport properties of the doped graphene leads. Here, we present an experiment probing those properties in the quantum Hall regime at low temperature and high magnetic field, showing that electronic phase coherence and transport chirality are preserved, despite the significant charge equilibration occurring at the edges of the leads. Our work yields a finer understanding of the properties of the doped graphene leads, allowing for improvements of the contact quality that can be applied to other two-dimensional materials.}
\end{abstract}

\maketitle

\begin{figure*}[ht]
\centering
\includegraphics[width=0.96\textwidth]{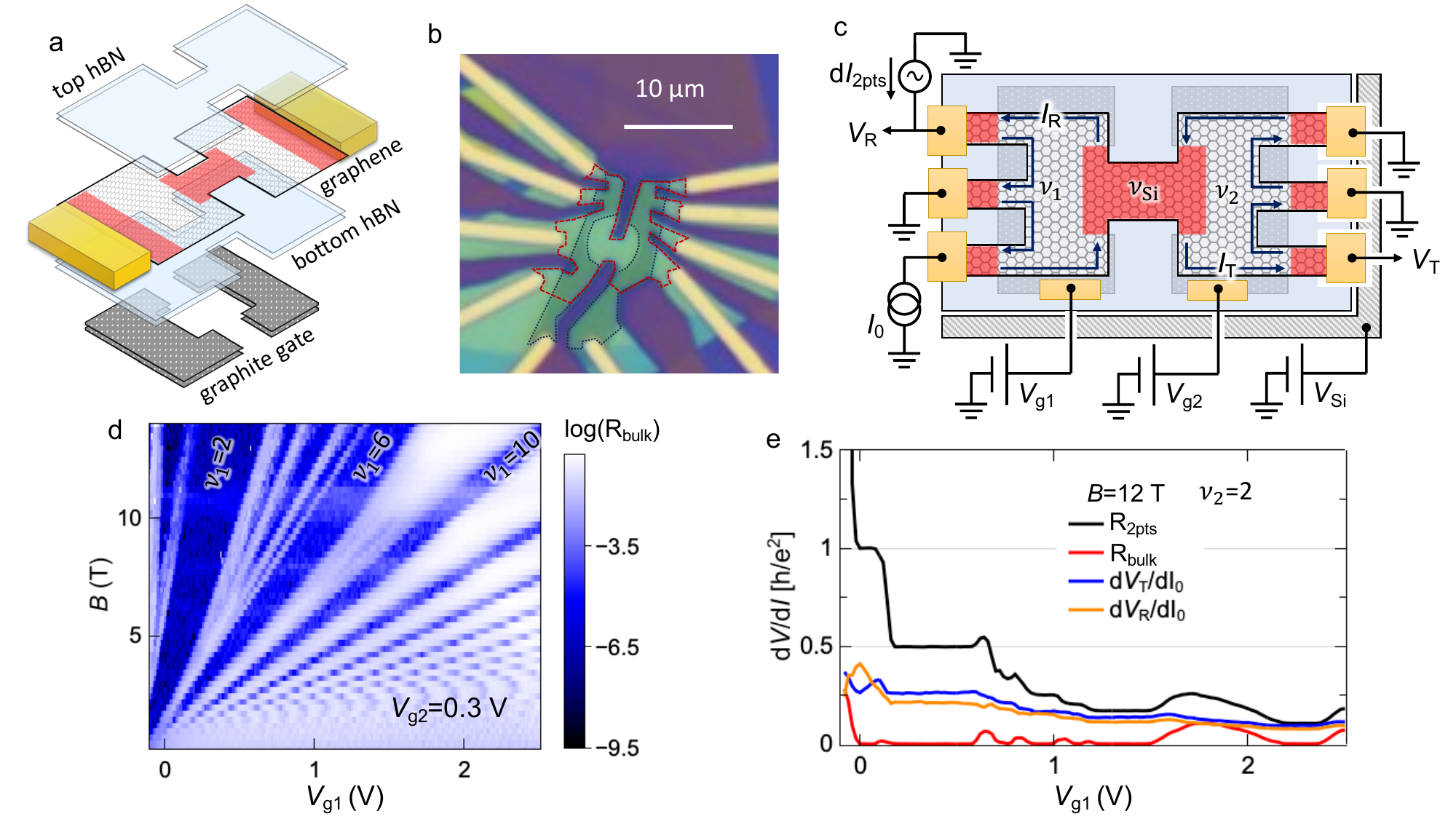}
\caption{\label{fig1-sample} Samples structure and typical high-field characterization. (a) Sketch of the heterostructure. The areas doped by the Si gate are highlighted in red. The yellow blocks represent the metallic contacts. (b) Optical micrograph of device A. The graphite gate is marked with the black dotted line, and the graphene flake is marked with the red dashed line. (c) Sketch of the devices, including the electrical connections. As in (a), the area of graphene doped by the Si gate is highlighted in red. The edge channels are depicted as dark blue arrows. (d) Landau fan diagram measurement of the transconductance $\dIR/\dIT$ (in log scale) of device A as a function of the magnetic field and graphite back gate voltage $\VgR$, at $T=10~$mK and $\VSi=40~$V. (e) Measurement of the differential resistances in device A as a function of the graphite back gate voltage $\VgR$, at $B=12~$T, $T=10~$mK, and $\VSi=60~$V. Black: 2-point resistance $R_\mathrm{2pts}$, red: bulk resistance $R_\mathrm{bulk}$, blue: transmitted transresistance $\dVT/\dIzero$, orange: reflected transresistance $\dVR/\dIzero$.
}
\end{figure*}

Electron transport and mesoscopic physics experiments in low-dimensional materials crucially depend on the quality of the contact between the metallic leads of the sample (connected to the measurement apparatus) and the material itself. In van der Waals heterostructures~\cite{Geim2013}, such as graphene encapsulated between two hexagonal boron nitride (hBN) crystals, this can be challenging, despite the recent development of one-dimensional edge contacts~\cite{Wang2013}. Recent experiments in ultra-clean graphene heterostructures, especially in the quantum Hall regime, have relied on a local electrostatic gating technique dubbed doped graphene leads~\cite{Maher2014} (or also doped graphene contacts~\cite{Ribeiro2019}),  whereby the carrier density in graphene close to the edge contacts is largely increased~\cite{Maher2014}, maximizing the electronic transmission between the metal and the lower density regions of the sample. This is usually done by incorporating in the heterostructure a graphite bottom gate of width smaller than that of the graphene flake, such that sizeable areas of the latter that extend outside of the area of the graphite gate. These areas are shaped into individual leads (of area typically $1~\mu\mathrm{m}^2$) connected to metallic edge contacts, and tuned to high carrier density (about $4-5\times10^{12}\mathrm{cm}^{-2}$)~\cite{Maher2014,Ribeiro2019} using a global electrostatic gate. This second gate is most commonly the doped silicon of the ($\mathrm{SiO}_2$/Si) substrate onto which the heterostructure is deposited~\cite{Maher2014}, but other variations of this method have been realized. For instance, the local graphite back gate can be replaced by a metallic gate prepatterned on the ($\mathrm{SiO}_2$/Si) substrate~\cite{Zibrov2017,Cao2018}, a ionic liquid deposited on top of the heterostructure can be used instead of the Si gate~\cite{Chuang2014}, or leads in transition metal dichalcogenides heterostructures can be doped with local top gates deposited on top of the sample~\cite{Shi2022}. Tuning the density of the graphene leads to high electron doping yields better transport measurement data (typically, more pronounced longitudinal resistance minima in the quantum Hall regime) when the graphite-gated part of the sample is also electron-doped (even at low densities), and vice-versa for hole doping~\cite{Maher2014,Ribeiro2019}. Oppositely, tuning the graphene leads and the graphite-gated region to opposite polarities degrades the signal, showing enhanced contact resistance, as well as fluctuations due to the p-n junction generated at the interface between the two graphene regions.

Despite being ubiquitous in modern graphene quantum Hall experiments, both in transport~\cite{Maher2014,Zeng2019,Ribeiro2019,Finney2022,Gul2022,Huang2022,Cohen2023} and capacitance measurements~\cite{Zibrov2017}, the actual transport properties of those graphene leads are often overlooked, as they are mostly considered a technical trick to enhance the contacts quality. Yet, they can host subtle physics, combining edge channels stemming from not-perfectly energy separated Landau levels, as well as several type of interfaces: graphene-metal, the etched edges of the graphene lead, and the interface between the graphene lead and the graphite-gated region. An important (if somewhat naive) question is how much "metal-like" do these graphene leads behave at high doping under large magnetic fields and at low temperature. Here we present measurements demonstrating residual signatures of hallmark quantum Hall transport phenomena in doped graphene leads: chirality, phase coherence, and charge equilibration.

\begin{figure*}[ht]
\centering
\includegraphics[width=0.94\textwidth]{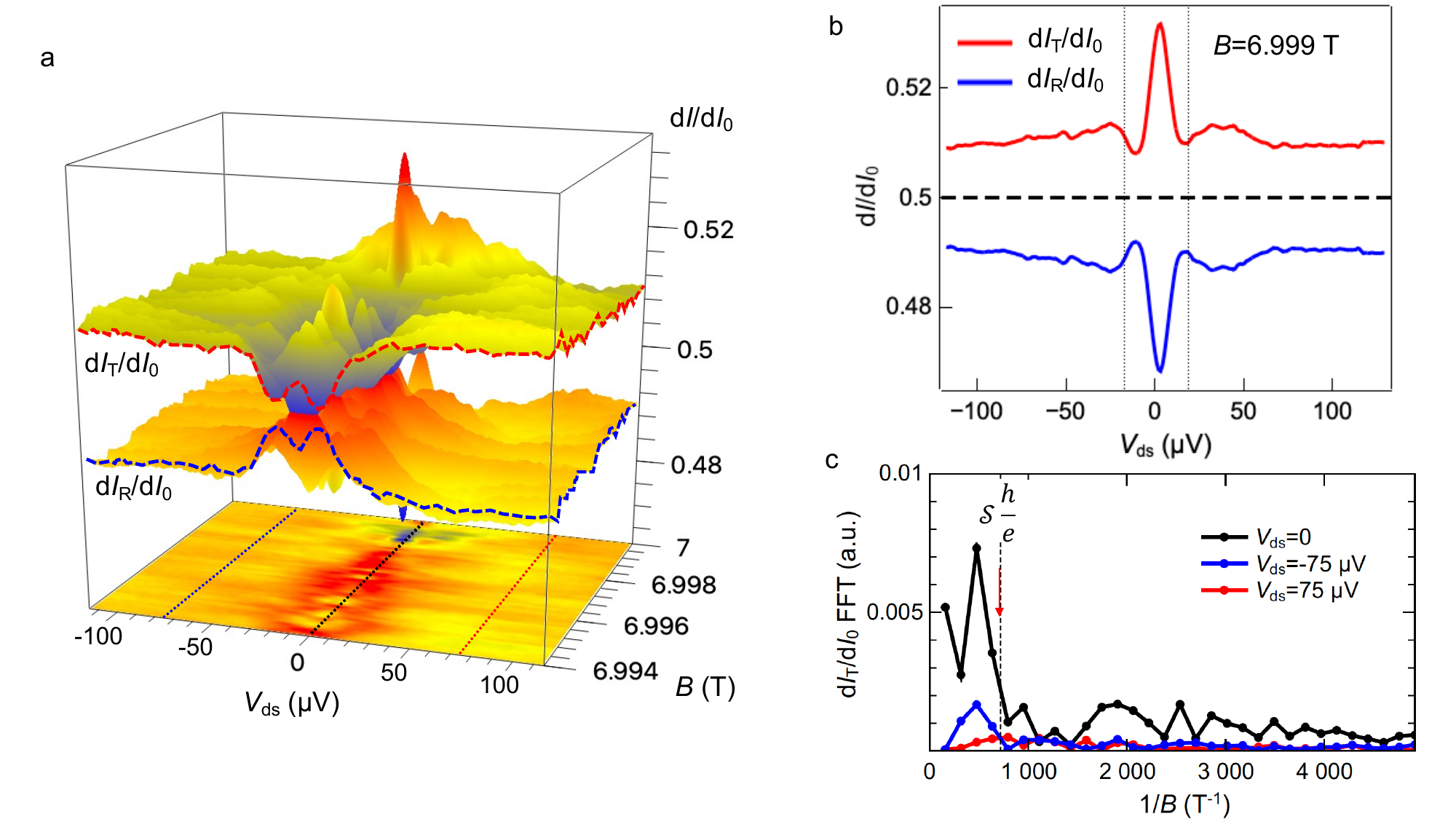}
\caption{\label{fig2-GvsBandVds} Transconductance measurements versus magnetic field and drain-source voltage in device A, for $\nu_1=\nu_2=2$. (a) Measured transmission $\dIT/\dIzero$ (marked with a red dashed line) and reflection $\dIR/\dIzero$ (marked with a blue dashed line) versus $\Vds$ (X-axis) and $B$ (Y-axis), for $\VSi=60~$V. The map in the bottom of the plot is a 2D projection of $\dIR/\dIzero$. (b) Linecuts of the data shown in (a) versus $\Vds$ for $B=6.999~$T, corresponding to the sharp peak (resp. dip) in $\dIT/\dIzero$ (resp. $\dIR/\dIzero$). The vertical dotted lines correspond to the Thouless energy along the perimeter of the doped graphene island (see text). (c) Fast Fourier transform of the linecuts of $\dIT/\dIzero$ versus $B$ at $\Vds=0$ (black), $\Vds=-75~\mu$V (blue), and $\Vds=75~\mu$V (red), corresponding to the dotted line in the 2D projection of (a). The vertical dashed line corresponds to the Aharonov-Bohm frequency expected from the area of the doped graphene island.
}
\end{figure*}

We performed the experiments on two different devices, the typical heterostructure of which is depicted in Figure~\ref{fig1-sample}a. It consists in a monolayer graphene flake encapsulated between two hBN crystals (typical thickness $20-40~$nm) and deposited on top of a few-layer graphite flake (typical thickness $5-10~$nm), prepatterned to incorporate a hole (typical width $\sim5~\mu$m). This heterostructure is deposited onto a ($\mathrm{SiO}_2$/Si) substrate, with a $\mathrm{SiO}_2$ thickness of $290~$nm. The graphene flake extends beyond the graphite flake, which we use a local back gate. This allows defining regions of the graphene flake that are not affected by the graphite gating effect, highlighted in red in Figure~\ref{fig1-sample}a: the regions extending outwards, connected to Cr/Au metallic electrodes using the edge contact technique \cite{Wang2013}, and the central region covering the hole, referred to in this manuscript as the graphene island. The doped Si substrate is used as a global back gate with which we tune the carrier density of those regions. At high carrier density, the outer regions thus act as standard graphene leads, similar to previous experiments~\cite{Maher2014,Zeng2019,Ribeiro2019,Gul2022,Huang2022,Cohen2023}. The heterostructure is processed to deposit the metallic edge contacts and etch the graphene flake so as to define two separate regions gated by the graphite flake, connected one to another via the graphene island. The graphite gate is also etched in two parts, allowing independently tuning the carrier densities of the two regions on either side of the graphene island.

An optical micrograph of one of the two devices presented in here (device A) is shown in Figure~\ref{fig1-sample}b, highlighting the edges of the graphene flake and the graphite back gate. The graphene island has a roughly circular shape, with area $\mathcal{S}\approx17~\mu\mathrm{m}^2$, and two types of edges: physical edges due to the etching, and gate-defined edges at the boundaries with the graphene regions doped by the graphite back-gate. Figure~\ref{fig1-sample}c details the principle of the experiment, and its experimental wiring for electron transport in the quantum Hall regime. The Si-gated regions are tuned using the gate voltage $\VSi$ to carrier density $n_\mathrm{Si}$, leading to filling factor $\nu_\mathrm{Si}=n_\mathrm{Si}h/eB$ under perpendicular magnetic field $B$ ($h$ is Planck's constant, and $e$ the electron charge). A current $I_0$ (including a small ac modulation $\dIzero$) is fed through the bottom left contact and its connected Si-doped graphene region, and is carried ballistically by the $\nu_1$ edge channels of the left graphite-doped region to the graphene island, tuned with graphite gate voltage $\VgR$. The portion of current transmitted to the right graphite-doped region (with filling factor $\nu_2$, tuned by the graphite gate voltage $\VgT$) is denoted $\IT$ and flows to a measurement contact (bottom right) where we measure the voltage drop $\VT=h/(\nu_1 e^2)\IT$ with respect to the cold ground connected to the contacts downstream. Conversely, the portion $\IR$ of current reflected to the left is measured using the upper left measurement contact, through the voltage drop $\VR=h/(\nu_2 e^2)\IR$. Additionally, a small ac current $\mathrm{d}I_\mathrm{2pts}$ is fed on that contact to measure the 2-point differential resistance to the ground $R_\mathrm{2pts}=\dVR/\mathrm{d}I_\mathrm{2pts}$, corresponding to the Hall resistance $h/\nu_1 e^2$ in the quantum Hall regime, as well as the bulk differential resistance $R_\mathrm{bulk}=\dVT/\mathrm{d}I_\mathrm{2pts}$, essentially acting as the equivalent of a longitudinal conductance~\cite{LeBreton2022}.

Our experiment consists in probing at low temperature ($10~$mK) the balance between $\IT$ and $\IR$ using low-frequency lock-in measurements~\cite{SM}, as well as their dependence with the gate voltages $\VgR$, $\VgT$, $\VSi$, the magnetic field $B$, and the drain-source dc voltage $\Vds=h/(\nu_1 e^2)I_0$. If the graphene island acts as a metallic lead at high doping $n_\mathrm{Si}$, the full equilibration of the edge channels entering and leaving the island should lead to balanced splitting of the currents  $I_{\mathrm{R}/\mathrm{T}}=I_0\times \nu_{1/2}/(\nu_1+\nu_2)$~\cite{Jezouin2013a,Srivastav2019}. This would appear as equal values of $\VR$ and $\VT$, given by the dc voltage developing at the island $\VR=\VT=I_0 (h/e^2)/(\nu_1+\nu_2)$. Figure~\ref{fig1-sample}d shows a Landau fan measurement of $R_\mathrm{bulk}$ in device A versus $B$ and $\VgR$, for $\VgT=0.3~$V. It displays large zones of vanishing $R_\mathrm{bulk}$, corresponding to gapped integer quantum Hall states in the graphite gated regions, including fully developed symmetry broken states in the $N=0$ and $N=1$ Landau levels. In these zones, the faint modulations of $R_\mathrm{bulk}$ with $B$, independent of $\VgR$, are attributed to Landau levels quantization in the graphite back gate~\cite{Wang2021,Zhu2021}. Transport properties across the graphene island are illustrated in Figure~\ref{fig1-sample}e, showing measurements of $\dVR/\dIzero$, $\dVT/\dIzero$, $R_\mathrm{2pts}$ and $R_\mathrm{bulk}$, versus $\VgR$, at $B=12~$T, $\VSi=60~$V ($n_\mathrm{Si}\approx4.3\times10^{12}~\mathrm{cm}^{-2}$), and $\VgT=0.3~$V, corresponding to $\nu_2=2$. While $R_\mathrm{2pts}$ shows well-developed plateaus at precise quantized values $h/(\nu_1 e^2)$, accompanied by zeroes of $R_\mathrm{bulk}$, the reflected and transmitted signals show plateaus at unequal values with a higher level of fluctuations, suggesting that the graphene island does not evenly split the current. Furthermore, the transmitted signal is generally larger than the reflected one, opposite to the behavior one would expect in presence of a non-negligible interface resistance between the graphene island and the graphite gated regions~\cite{Srivastav2019}. This is clearly visible at $\nu_1=\nu_2=2$, which is the configuration we focus on throughout the rest of this letter.

\begin{figure}[ht]
\centering
\includegraphics[width=0.47\textwidth]{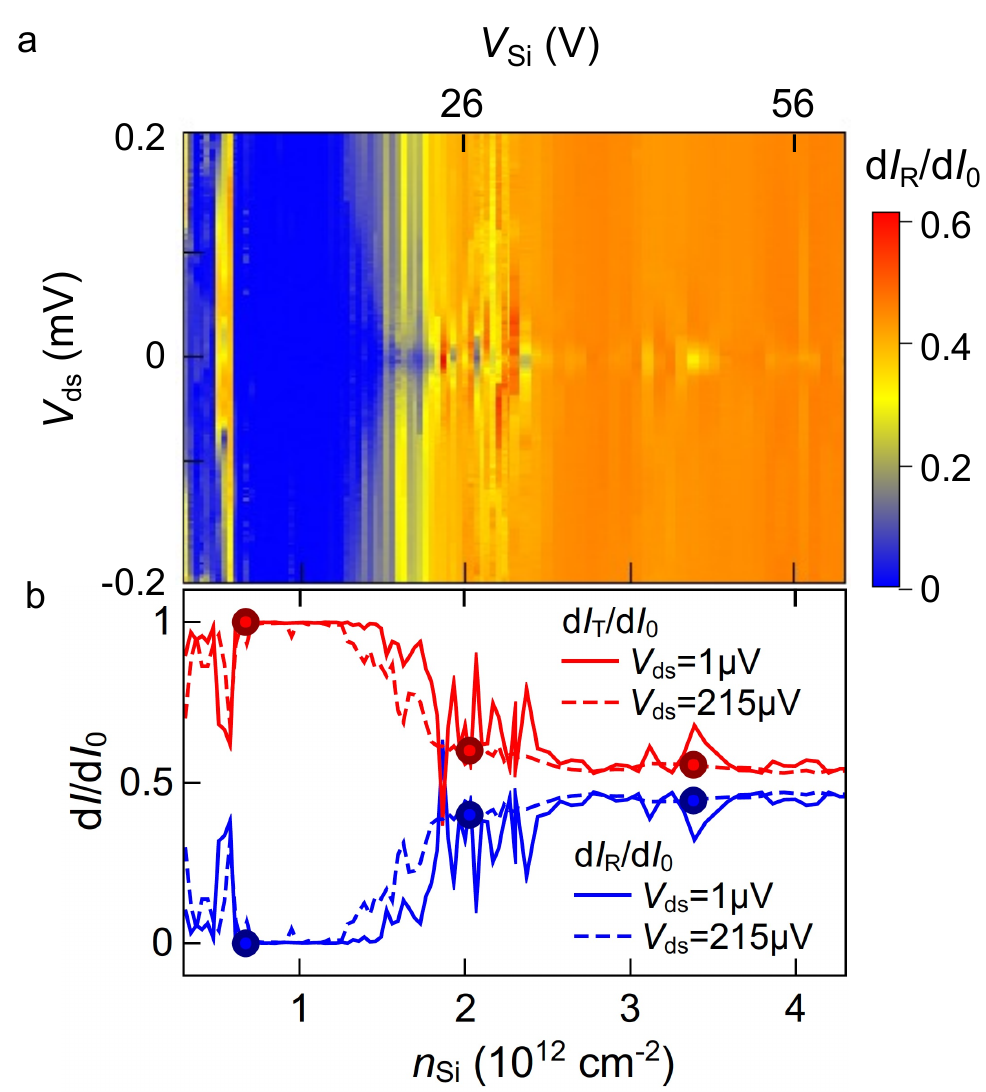}
\caption{\label{fig3-GvsnSiandVds} Transconductance measurements versus Si gate voltage (tuning the density of the graphene island) and drain-source voltage in device A, at $B=14~$T, for $\nu_1=\nu_2=2$. (a) Reflection $\dIR/\dIzero$ plotted versus $\Vds$ and $n_\mathrm{Si}$. (b) Linecuts of $\dIR/\dIzero$ (blue, corresponding to the data shown in (a)) and of $\dIT/\dIzero$ (red) versus $n_\mathrm{Si}$, for $\Vds=1~\mu$V (full lines) and $\Vds=215~\mu$V (dashed lines). Circles: calculations of the transmission (red) and reflection (blue) in presence of full incoherent mixing on both sides of the doped graphene island (see text).
}
\end{figure}

Figure~\ref{fig2-GvsBandVds}a shows a measurement of the transmitted and reflected signals $\dIT/\dIzero$ and $\dIR/\dIzero$ in device A at $\nu_1=\nu_2=2$ versus drain-source voltage and magnetic field, for $\VSi=60~$V. The signals are unequal and symmetric, with seemingly random oscillatory features both in $\Vds$ and $B$. These oscillations are dampened at finite bias over a typical scale of $\sim20~\mu$V. This is exemplified by the line cut at constant $B$ plotted in Figure~\ref{fig2-GvsBandVds}b: $\dIT/\dIzero$ (red) and $\dIR/\dIzero$ (blue) show opposite oscillations as $|\Vds|$ increases, and saturate at $|\Vds|>50~\mu$V constant values $\dIT/\dIzero>\dIR/\dIzero$, again in contradiction of what one would expect from an interface resistance. The typical scale of the oscillations and their decay is comparable with the Thouless energy $h v/L$ (vertical dotted lines in Figure~\ref{fig2-GvsBandVds}b) associated with the time of flight $L/v$ of an excitation travelling around the perimeter of the graphene island $L\approx21~\mu$m (extracted from the optical micrograph of Figure~\ref{fig1-sample}b) and a typical excitation velocity in graphene quantum Hall edge channels $v=10^5~$m$/$s~\cite{Deprez2021,Ronen2021}. Similarly, fast Fourier transform analysis of the linecuts at fixed $\Vds$, plotted in Figure~\ref{fig2-GvsBandVds}c, show a dominant component at $(1.4~\mathrm{mT})^{-1}$, slightly lower than the Aharonov-Bohm frequency $\mathcal{S}\times h/e$ predicted from the area of the graphene island. This suggests that the observed oscillations are related to quantum Hall Fabry-Perot type~\cite{Chamon1997} electronic interferences in the graphene island, which we discuss in more details in the last part of this letter.

\begin{figure}[ht]
\centering
\includegraphics[width=0.42\textwidth]{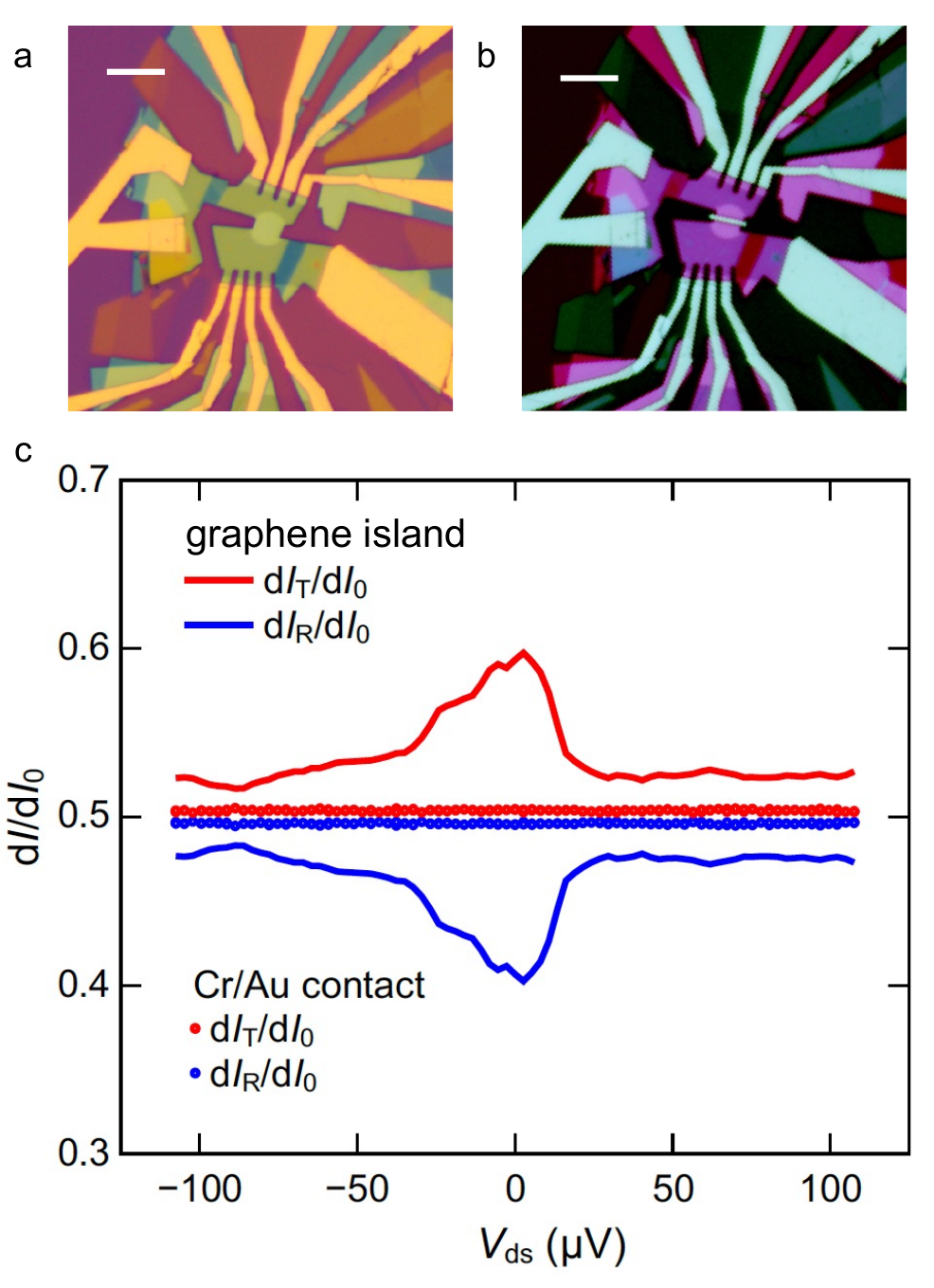}
\caption{\label{fig4-GvsVds-Rst07} Transconductance measurements in device B with a doped graphene island and a metallic contact, for $\nu_1=\nu_2=2$. (a,b) Optical micrographs of device B before (a) and after (b) deposition of a Cr/Au metallic electrode interrupting the doped graphene island. Scale bar: $10~\mu$m. (c) Measurements of $\dIT/\dIzero$ (red) and $\dIR/\dIzero$ (blue) versus $\Vds$. Lines are data measured at $B=11~$T and $\VSi=60~$V before deposition of the metallic electrode; symbols are data measured $B=14~$T and $\VSi=0~$V (see text) after deposition.
}
\end{figure}

\begin{figure*}[ht]
\centering
\includegraphics[width=0.94\textwidth]{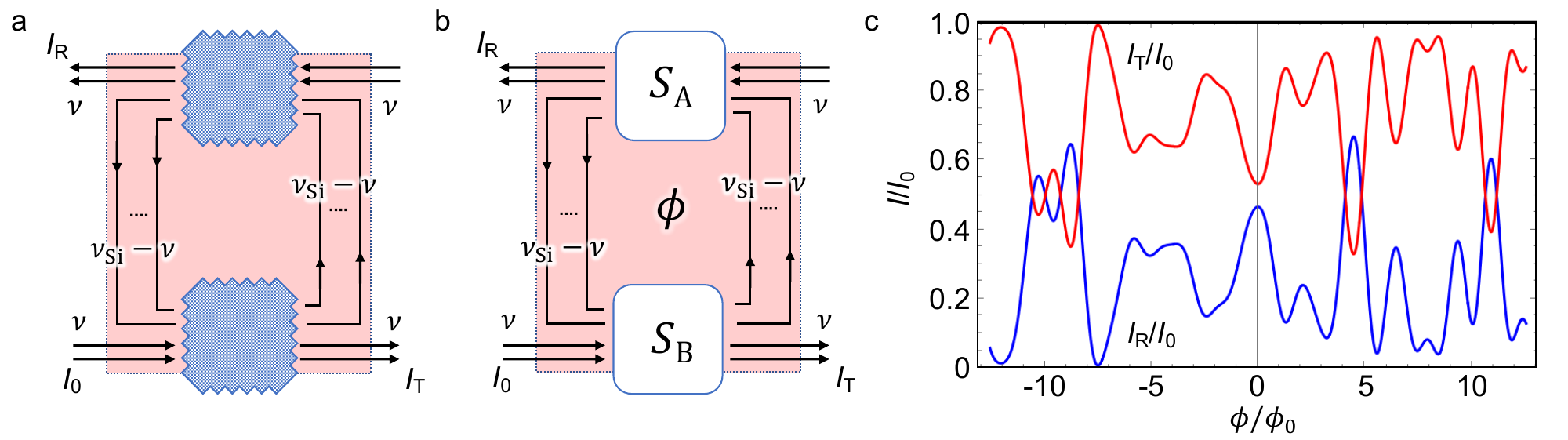}
\caption{\label{fig5-models} Transport models. (a) Sketch of the non-phase coherent model including two voltage probes (blue serrated squares) equilibrating the edge channels (black arrows) on each edge of the doped region, highlighted in red. (b) Sketch of the phase coherent model where the voltage probes of (a) are replaced by two independent $\nu_\mathrm{Si}\times\nu_\mathrm{Si}$ scattering matrices (blue squares) that mix the edge channels together. A magnetic flux $\phi$ is enclosed in the largest loop formed by the two outermost edge channels in the doped island. (c) Typical result of the phase coherent model, showing the variation of the transmission $\IT/I_0$ (red) and the reflection $\IR/I_0$ (blue) versus magnetic flux (see text for details).
}
\end{figure*}

Figure~\ref{fig3-GvsnSiandVds}a shows the dependence of the reflected signal in device A as a function of the drain-source voltage and the Si gate voltage. At low $\VSi$, the reflected signal is zero, then becomes finite with low-bias oscillatory features similar to that of Figure~\ref{fig2-GvsBandVds}. These features persist up to large $\VSi$; notably, at large bias, $\dIR/\dIzero$ saturates to values systematically smaller than $0.5$. This is demonstrated in the corresponding linecuts of $\dIR/\dIzero$ and $\dIT/\dIzero$ plotted versus $n_\mathrm{Si}$ in Figure~\ref{fig3-GvsnSiandVds}b. The vanishing $\dIR/\dIzero$ at $\VSi<20~$V ($n_\mathrm{Si}<1.5\times10^{12}~\mathrm{cm}^{-2}$) is accompanied by a quantized value $\dIT/\dIzero=1$, which corresponds to the formation of the $\nu_\mathrm{Si}=2$ quantum Hall plateau in the island, thus fully transmitting the two edge channels. Increasing $n_\mathrm{Si}$ leads to the symmetric increase (resp. decrease) of $\dIR/\dIzero$ (resp. $\dIT/\dIzero$) with symmetric oscillatory features that dampen and large bias. Figure~\ref{fig3-GvsnSiandVds}b further highlights the saturation of both signals at $n_\mathrm{Si}>2.5\times10^{12}~\mathrm{cm}^{-2}$, never achieving equal current splitting (note that we were not able to apply gate voltages larger than $60~$V due to leaks occurring in our cryostat wiring).

The central role of the graphene island in these observations was checked with device B, where we compared the reflected and transmitted signals at low temperature and high magnetic field before and after depositing a  metal electrode interrupting the graphene island. This is shown in Figure~\ref{fig4-GvsVds-Rst07}, displaying the optical micrographs of device B without (Figure~\ref{fig4-GvsVds-Rst07}a) and with (Figure~\ref{fig4-GvsVds-Rst07}b) a $6\times0.8~\mu\mathrm{m}^2$ Cr/Au edge contact. This yields drastic changes in transport, illustrated in the $\Vds$ dependence of $\dIR/\dIzero$ (blue) and $\dIT/\dIzero$ (red): in presence of the metallic contact, the oscillatory behavior completely vanishes and both signals are much closer to $0.5$ at $\nu_1=\nu_2=2$. The transmitted signal is still slightly higher, by a few percent, than the reflected one. This may be due to calibration errors; in particular, during this second cooldown of device 2, $\VSi$ was kept at zero due to a leakage, which can increase the contact resistance of the current feed and cold ground.

Our results, and in particular the fact that the transmitted signal is consistently higher than the reflected one, suggest that chirality is preserved in the graphene island. We use two models based on a Landauer-Büttiker formalism~\cite{Moskalets2011}, shown in Figure~\ref{fig5-models}, to understand our results. Both models rely on the edge channels being preserved at the interfaces between the graphene island and the graphite-gated regions, while edge channels mixing occurs along the etched edges of the island. This is justified by the higher degree of disorder at an etch-defined edge, having been shown to lead to edge channel mixing~\cite{Amet2014}. The mixing is modelled by two different mechanisms, with $\nu_1=\nu_2=\nu$ for simplicity. In Figure~\ref{fig5-models}a, it is described by voltage probes~\cite{Buttiker1988,deJong1996,Forster2007} that fully equilibrate all $\nu_\mathrm{Si}$ edge channels flowing along the etched edge of the graphene island, suppressing phase coherence. Calculating the current balance at each of the voltage probes~\cite{SM} leads to $\IR=I_0\times(\nu_\mathrm{Si}-\nu)/(2\nu_\mathrm{Si}-\nu)$ and $\IT=I_0\times(\nu_\mathrm{Si})/(2\nu_\mathrm{Si}-\nu)$. The symbols in Figure~\ref{fig3-GvsnSiandVds}b are given by these expressions, for $\nu=2$ and the values of $\VSi$ and $n_\mathrm{Si}$ corresponding to $\nu_\mathrm{Si}=\{2,6,10\}$, showing an excellent agreement with both the data values and their trend. This calculation shows that even for very large filling factors $\nu_\mathrm{Si}\sim50$, current splitting remains imperfect~\cite{SM}. Despite this good quantitative agreement, the model fails to capture the low-bias oscillations, as phase coherence is suppressed by the voltage probes, preventing electronic interferences around the perimeter of the island. In the second model, shown in Figure~\ref{fig5-models}b, the voltage probes are replaced by two scattering matrices $\mathcal{S}_\mathrm{A}$ and $\mathcal{S}_\mathrm{B}$ that describe coherent point-like mixing between all $\nu_\mathrm{Si}$ edge channels flowing along the island's etched edges. The different Fabry-Perot interferometers created by the $\nu_\mathrm{Si}-\nu$ channels flowing along the gate-defined interfaces lead to the accumulation of different Aharonov-Bohm phases due to the distance between each edge channels, which gives rise to seemingly random interferences. This is illustrated in Figure~\ref{fig5-models}c, which plots the result of numerical calculations with $\nu_\mathrm{Si}=5$, $\nu=1$, and taking into account up to three round trips around the graphene island. $\mathcal{S}_\mathrm{A}$ and $\mathcal{S}_\mathrm{B}$ are generated randomly, and the Aharonov-Bohm phases corresponding to the $4$ edges channels flowing in the island are arbitrarily set with respect to the one of the innermost channel to $\{1.65\phi,1.6\phi,1.5\phi,\phi\}$. The reflected (blue) and transmitted (red) currents show symmetric and seemingly random oscillations, with $I_\mathrm{T}$ generally above $I_\mathrm{R}$, again emphasizing the chirality of transport in the island. Note that the amplitude of the oscillations is of order unity, as no decoherence mechanism is included in this model. 

The comparison of our experiment and the two models thus shows that the transport properties of doped graphene leads are far from that of a simple metal: despite the large carrier density, chirality is preserved, as is phase coherence along the edge channels flowing at the interface between the graphene lead and the graphite-gated regions. Oppositely, along the etched edges of the graphene lead, significant edge channel equilibration occurs; thus, in order to minimize the contact resistance, one should increase the number of etched edges by dividing the graphene lead in several parallel narrow strips, as was recently done in experiments on graphene quantum point contacts~\cite{Cohen2023a} and Fabry-Perot interferometers~\cite{Samuelson2024} in the quantum Hall regime.

\textbf{Acknowledgements}

 This work was funded by the ERC (ERC-2018-STG \textit{QUAHQ}), by the “Investissements d’Avenir” LabEx PALM (ANR-10-LABX-0039-PALM), and by the Region Ile de France through the DIM SIRTEQ. The authors warmly thank R. Ribeiro-Palau for enlightening discussions, and P. Jacques for technical support.


\providecommand{\latin}[1]{#1}
\makeatletter
\providecommand{\doi}
  {\begingroup\let\do\@makeother\dospecials
  \catcode`\{=1 \catcode`\}=2 \doi@aux}
\providecommand{\doi@aux}[1]{\endgroup\texttt{#1}}
\makeatother
\providecommand*\mcitethebibliography{\thebibliography}
\csname @ifundefined\endcsname{endmcitethebibliography}  {\let\endmcitethebibliography\endthebibliography}{}

\end{document}


\date{\today}

\maketitle

\begin{figure*}[ht]
\centering
\includegraphics[width=0.96\textwidth]{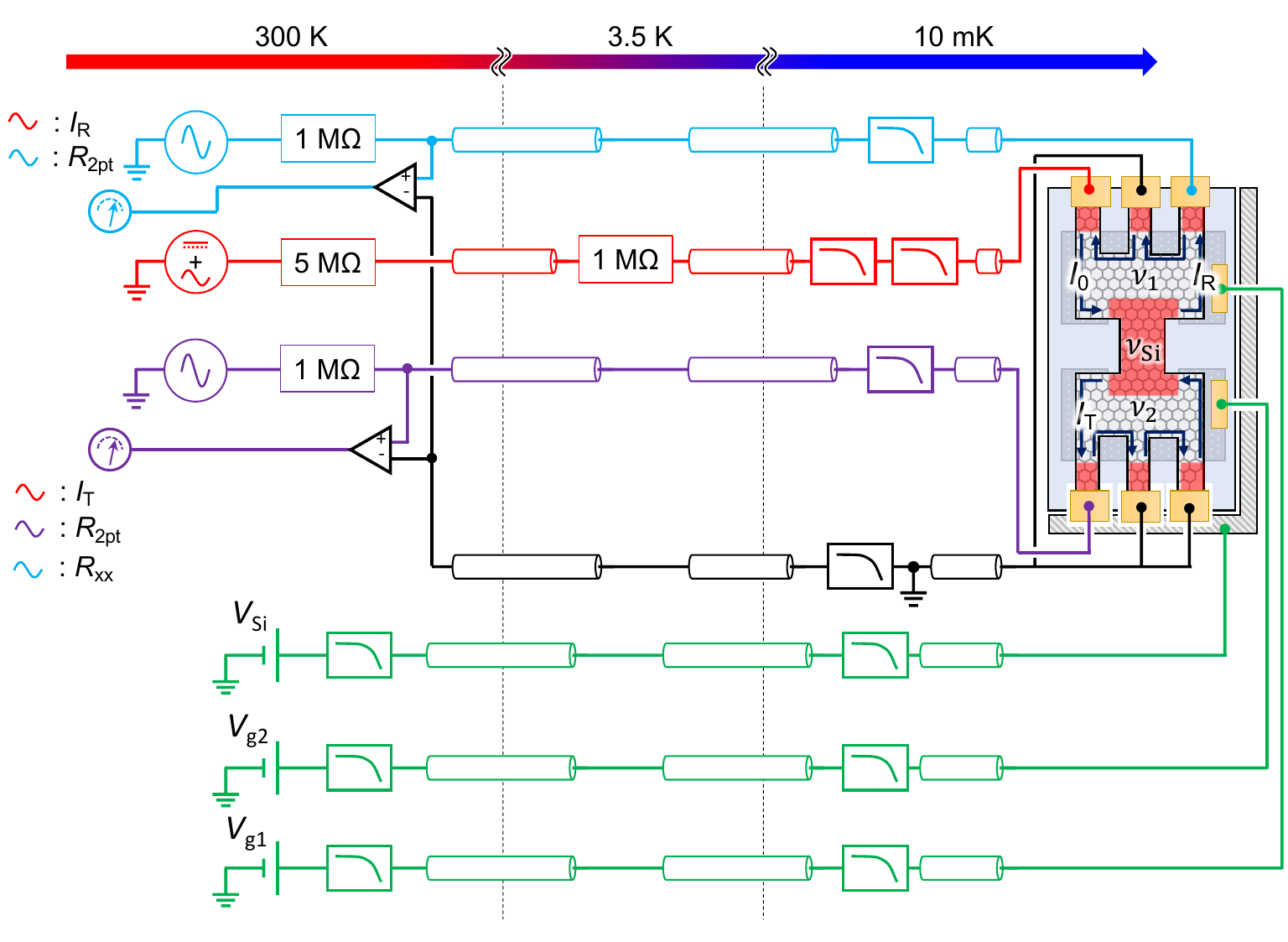}
\caption{\label{figsup-conductances} Layout of the wiring for the conductance measurements. Lines are color-coded (blue: R-side conductance measurements; green: Si and graphite gates; red: $I_0$ current feed; purple: T-side conductance measurements; black: cold ground).
}
\end{figure*}

\newpage
\section{Samples fabrication}

The samples were fabricated by first picking up a graphite flake using the a PPC stamp, then depositing it onto a ($\mathrm{SiO}_2$/Si) substrate with pre-patterned leads and bonding pads. The hole in the graphite flake was defined using electron-beam lithography, and etched with an oxygen plasma. The PMMA resist was removed with aceton and isopropanol, after which the flake was annealed in vacuum at $400~^\circ$C for 45 minutes. A hBN / graphene / hBN stack was then deposited using a PPC stamp on the patterned graphite flake, annealed in vacuum, then processed using standard fabrication techniques for hBN etching (CHF$_3$/O$_2$ plasma) and Cr/Au metallic electrodes deposition. The area and perimeter of the doped islands, measured with an optical microscope, are $16~\mu$m$^2$ and $21~\mu$m for device A, and $27.5~\mu$m$^2$ and $21.7~\mu$m for device B. The metallic contact added before the second cooldown of device B was fabricated in the same manner, after removing the bonding wires of the sample and taking it out of its sample holder.

\section{Experimental setup}

A detailed description of the conductance measurements is shown in Supplementary Fig.~\ref{figsup-conductances}. The measurements were performed using lock-in techniques at low frequency, below 100~Hz. All lines, including current feed (red in Fig.~\ref{figsup-conductances}) and gates (green in Fig.~\ref{figsup-conductances}) are filtered at the mixing chamber stage of our dilution refigerator. All measurements are performed using differential amplifiers (CELIANS EPC-1B) referenced to the cold ground (black in Fig.~\ref{figsup-conductances}) The latter is directly connected (both electrically and thermally) to the mixing chamber stage. The current feed line includes a $1~$M$\Omega$ series bias resistor thermally anchored to the $3.5~$K stage of our dilution refrigerator.

The effect of the filters (both in terms of series resistance and capacitive cutoff) are taken into account in our data. In particular, in the data corresponding to main text Figures 2 and 3, the transmitted and reflected signals are normalized by their sum: 

\begin{eqnarray}
\left[\mathrm{d}I_{\mathrm{T}/\mathrm{R}}/\dIzero\right]_\mathrm{norm}=\frac{\left[\mathrm{d}I_{\mathrm{T}/\mathrm{R}}/\dIzero\right]_\mathrm{raw}}{\left[\mathrm{d}I_{\mathrm{T}}/\dIzero\right]_\mathrm{raw}+\left[\mathrm{d}I_{\mathrm{R}}/\dIzero\right]_\mathrm{raw}},
    \label{eq:normalization}
\end{eqnarray}

in order to take into account the capacitive cutoff on the current feed line. We show in Supplementary Figure~\ref{figsub-normalization} the comparison between the raw and normalized signals, as well as the sum of the raw signals $\left[\mathrm{d}I_{\mathrm{T}}/\dIzero\right]_\mathrm{raw}+\left[\mathrm{d}I_{\mathrm{R}}/\dIzero\right]_\mathrm{raw}$, corresponding to the data of main text Figure 2. The sum of the raw signals is constant (corresponding to current conservation in the sample), but about $3~\%$ lower than its expected value because of the capacitive cutoff in the lines. Correspondingly, the raw reflected and transmitted signals (thin blue and red diamonds) show a symmetry with respect to a value $3~\%$ lower than half-transmission (crucially, the raw reflected signal is still lower than the raw transmitted one). Normalizing the raw signals by their sum yields the data plotted as thick circles, shifted upwards by about $3~\%$ and symmetric with respect to $0.5$.

\begin{figure}[ht!]
\centering
\includegraphics[width=0.57\textwidth]{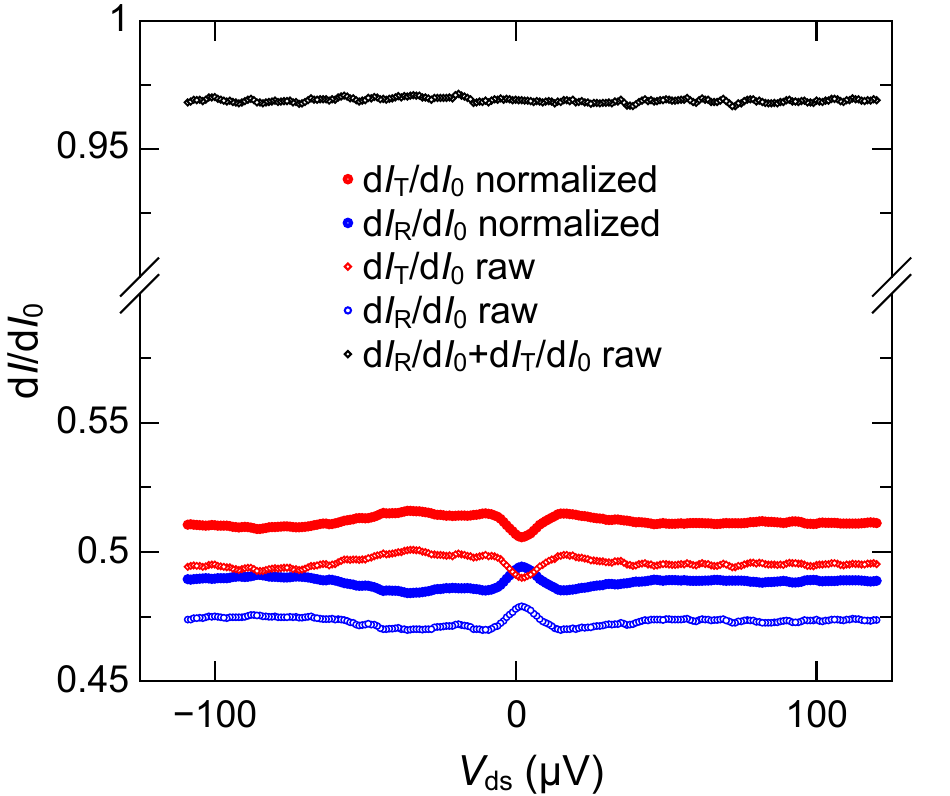}
\caption{\label{figsub-normalization} Transmitted and reflected signals measured versus $\Vds$ at $B=7~$T in device A, corresponding to the data shown in main text Figure 2. Thick circles: normalized data (shown in the main text), thin diamonds: raw data. Black: sum of the raw transmitted and reflected signals.
}
\end{figure}

\section{Transport models}
\subsection{Voltage probes}

\begin{figure}[ht!]
\centering
\includegraphics[width=0.57\textwidth]{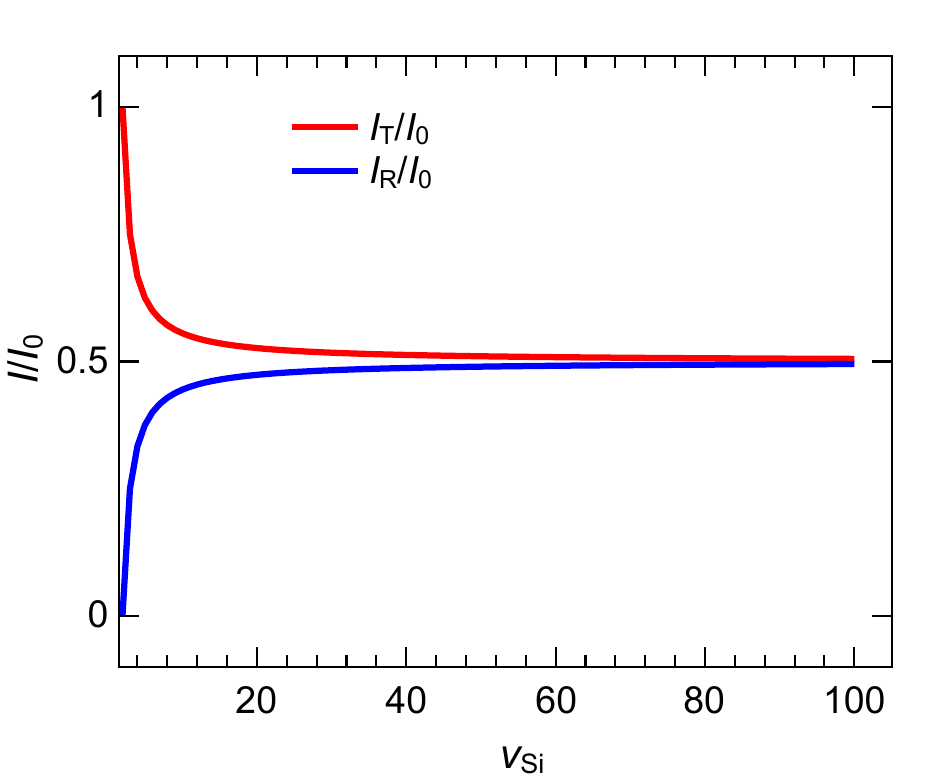}
\caption{\label{figsub-simuvoltageprobe} Calculated transmitted and reflected currents versus $\nu_\mathrm{Si}$ with the voltage probe model, for $\nu=2$.
}
\end{figure}

In the voltage probes model, the equilibration of the edge channels along the etched edges of the graphene island is assumed to be perfect. Voltage probes emulating this perfect equilibration are thus introduced in both upper and lower edges, and the transmitted and reflected currents are calculated through the current balance on each probe, according to the sketch depicted in main text Figure 5:

\begin{eqnarray}
    I_0+(\nu_\mathrm{Si}-\nu)G_0 V_\mathrm{A}=\nu_\mathrm{Si}G_0 V_\mathrm{B}\\
   (\nu_\mathrm{Si}-\nu)G_0 V_\mathrm{B}=\nu_\mathrm{Si}G_0 V_\mathrm{A},
    \label{eq:currbalvoltageprobe1}
\end{eqnarray}

where $V_\mathrm{A}$ and $V_\mathrm{B}$ are the voltages developing in (respectively) the probe on the upper and lower edge, and $G_0=e^2/h$ is the electrical conductance quantum. In this calculation, $\nu_1$ and $\nu_2$ are taken equal for simplicity: $\nu_1=\nu_2=\nu$. Solving these equations yields the reflected and transmitted currents:

\begin{eqnarray}
    \IT=\nu G_0 V_\mathrm{B}=I_0\times\frac{\nu_\mathrm{Si}}{2\nu_\mathrm{Si}-\nu}\\
    \IR=\nu G_0 V_\mathrm{A}=I_0\times\frac{\nu_\mathrm{Si}-\nu}{2\nu_\mathrm{Si}-\nu}
    \label{eq:currbalvoltageprobe2}
\end{eqnarray}

The points plotted in main text Figure 3b are correspond to these expressions with $\nu=2$ and $\nu_\mathrm{Si}=\{2,4,6\}$. Figure~\ref{figsub-simuvoltageprobe} shows the result of the calculation for fillong factors up to $\nu_\mathrm{Si}=100$, showing about a percent of deviation to perfect current splitting.

\subsection{Phase coherent model}

In the phase coherent model, partial mixing at the edges is described by two $\nu_\mathrm{Si}\times\nu_\mathrm{Si}$ scattering matrices $\mathcal{S}_\mathrm{A}^{\alpha,\beta}$ and $\mathcal{S}_\mathrm{B}^{\alpha,\beta}$, with $\alpha=\beta=1$ corresponding to the edge channel flowing in the graphite gated regions, the filling factors of which are set to $\nu_1=\nu_2=1$ for simplicity. An electron incoming from the current feed contact is thus reflected by the island with a probability:

\begin{eqnarray}
    \mathcal{R}(\phi)=|\sum_n t_n(\phi)|^2,
    \label{eq:currbalvoltageprobe1}
\end{eqnarray}

where $t_n(\phi)$ is the amplitude acquired after $n$ round trips around the island, and involves the scattering of electrons between all $\nu_\mathrm{Si}-\nu=\nu_\mathrm{Si}-1$ edge channels flowing around the island, shown below for up to three round trips, as used in our calculations:

\begin{eqnarray}
    t_0(\phi)=\sum_{j=2}^{\nu_\mathrm{Si}} \mathcal{S}_\mathrm{A}^{1,j}\mathcal{S}_\mathrm{B}^{j,1}\times e^{i\phi_j/2}
    \\
    t_1(\phi)=\sum_{j,k,k'=2}^{\nu_\mathrm{Si}} \mathcal{S}_\mathrm{A}^{1,j}\mathcal{S}_\mathrm{B}^{j,k}\mathcal{S}_\mathrm{A}^{k,k'}\mathcal{S}_\mathrm{B}^{k',1}\times e^{i(\phi_j+\phi_k+\phi_{k'})/2}
    \\
    t_2(\phi)=\sum_{j,k,k',l,l'=2}^{\nu_\mathrm{Si}} \mathcal{S}_\mathrm{A}^{1,j}\mathcal{S}_\mathrm{B}^{j,k}\mathcal{S}_\mathrm{A}^{k,k'}\mathcal{S}_\mathrm{B}^{k',l}\mathcal{S}_\mathrm{A}^{l,l'}\mathcal{S}_\mathrm{B}^{l',1}\times e^{i(\phi_j+\phi_k+\phi_{k'}+\phi_l+\phi_{l'})/2}
    \\
    t_3(\phi)=\sum_{j,k,k',l,l',m,m'=2}^{\nu_\mathrm{Si}} \mathcal{S}_\mathrm{A}^{1,j}\mathcal{S}_\mathrm{B}^{j,k}\mathcal{S}_\mathrm{A}^{k,k'}\mathcal{S}_\mathrm{B}^{k',l}\mathcal{S}_\mathrm{A}^{l,l'}\mathcal{S}_\mathrm{B}^{l',m}\mathcal{S}_\mathrm{A}^{m,m'}\mathcal{S}_\mathrm{B}^{m',1}\times e^{i(\phi_j+\phi_k+\phi_{k'}+\phi_l+\phi_{l'}+\phi_{m}+\phi_{m'})/2},
    \label{eq:currbalvoltageprobe2}
\end{eqnarray}

with $\phi_k/2$ the Aharonov-Bohm phase acquired  by an electron in the $k-$th edge channel in one half-round trip. We restrict our numerical calculations to maximum three round trips, with $\nu_\mathrm{Si}=5$ and $\nu_1=\nu_2=1$. The scattering matrices $\mathcal{S}_\mathrm{A}^{\alpha,\beta}$ and $\mathcal{S}_\mathrm{B}^{\alpha,\beta}$ are generated randomly, and we arbitrarily parameterize the four phases by $\{1.65\phi,1.6\phi,1.5\phi,\phi\}$.

\begin{figure}[ht]
\centering
\includegraphics[width=0.99\textwidth]{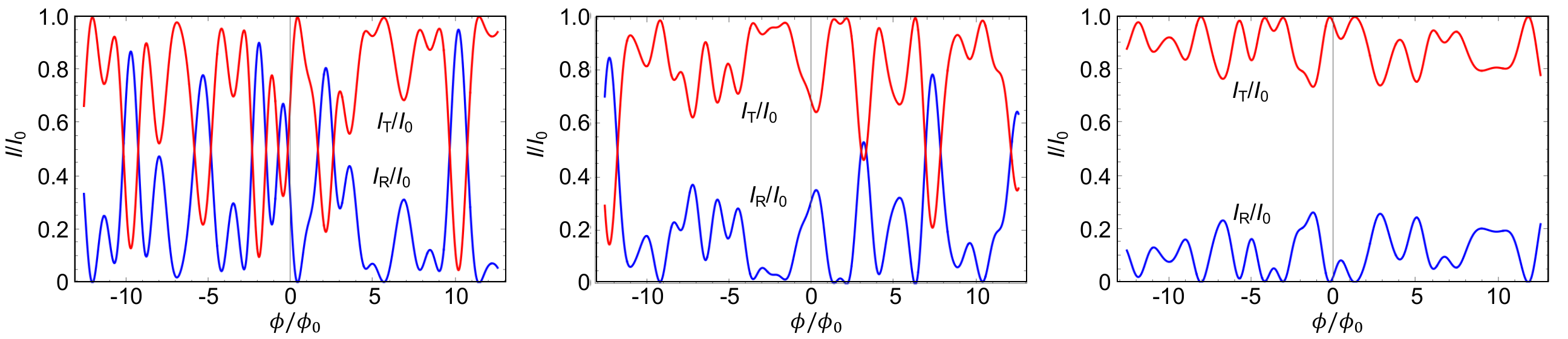}
\caption{\label{figsub-simus} Results of the phase coherent model, showing the variation of the transmission $\IT/I_0$ (red) and the reflection $\IR/I_0$ (blue) versus magnetic flux, for three sets of randomly generated scattering matrices (left: $|\mathcal{S}_\mathrm{B}^{1,1}|^2\approx0.37$; center: $|\mathcal{S}_\mathrm{B}^{1,1}|^2\approx.54$; right: $|\mathcal{S}_\mathrm{B}^{1,1}|^2\approx0.77$.
}
\end{figure}

Figure~\ref{figsub-simus} shows additional results of the model, for the same phase parameters, but different randomly generated scattering matrices. The baseline and amplitude of the oscillations appear to be loosely correlated to the coefficient $|\mathcal{S}_\mathrm{B}^{1,1}|^2$, which corresponds to the probability for an electron stemming from the current feed contact on the left side of the island to be directly transmitted to the right side of the island without entering the interferometer. The higher this probability, the larger the discrepancy between transmitted and reflected signals, and the smaller the oscillations. In Figure 5c of the main text, $|\mathcal{S}_\mathrm{B}^{1,1}|^2\approx0.45$.

\newpage
\section{Additional data - device A}



\begin{figure}[ht!]
\centering
\includegraphics[width=0.57\textwidth]{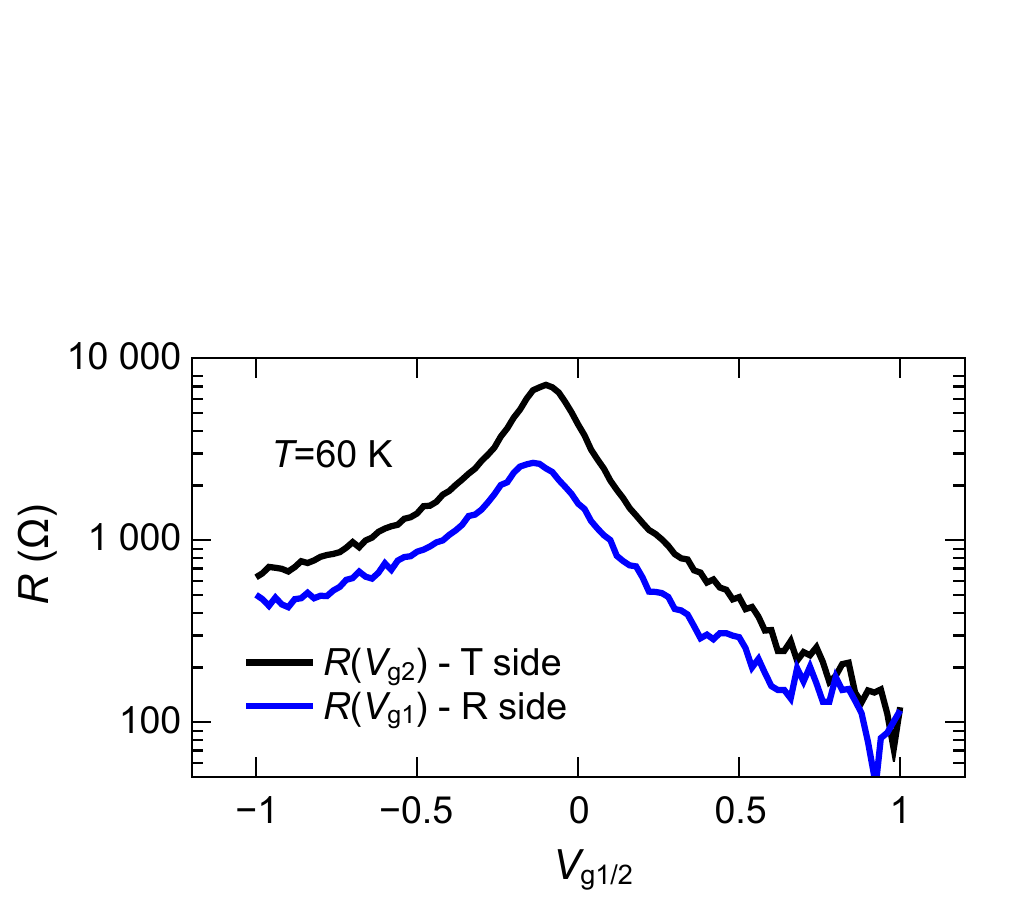}
\caption{\label{figsub-RvsVgDeviceA} Measurement of the two point resistance in device A versus graphite gate voltages, at $B=0~$T, $T=60~$K, and $\VSi=0~$V.
}
\end{figure}

\begin{figure}[ht!]
\centering
\includegraphics[width=0.67\textwidth]{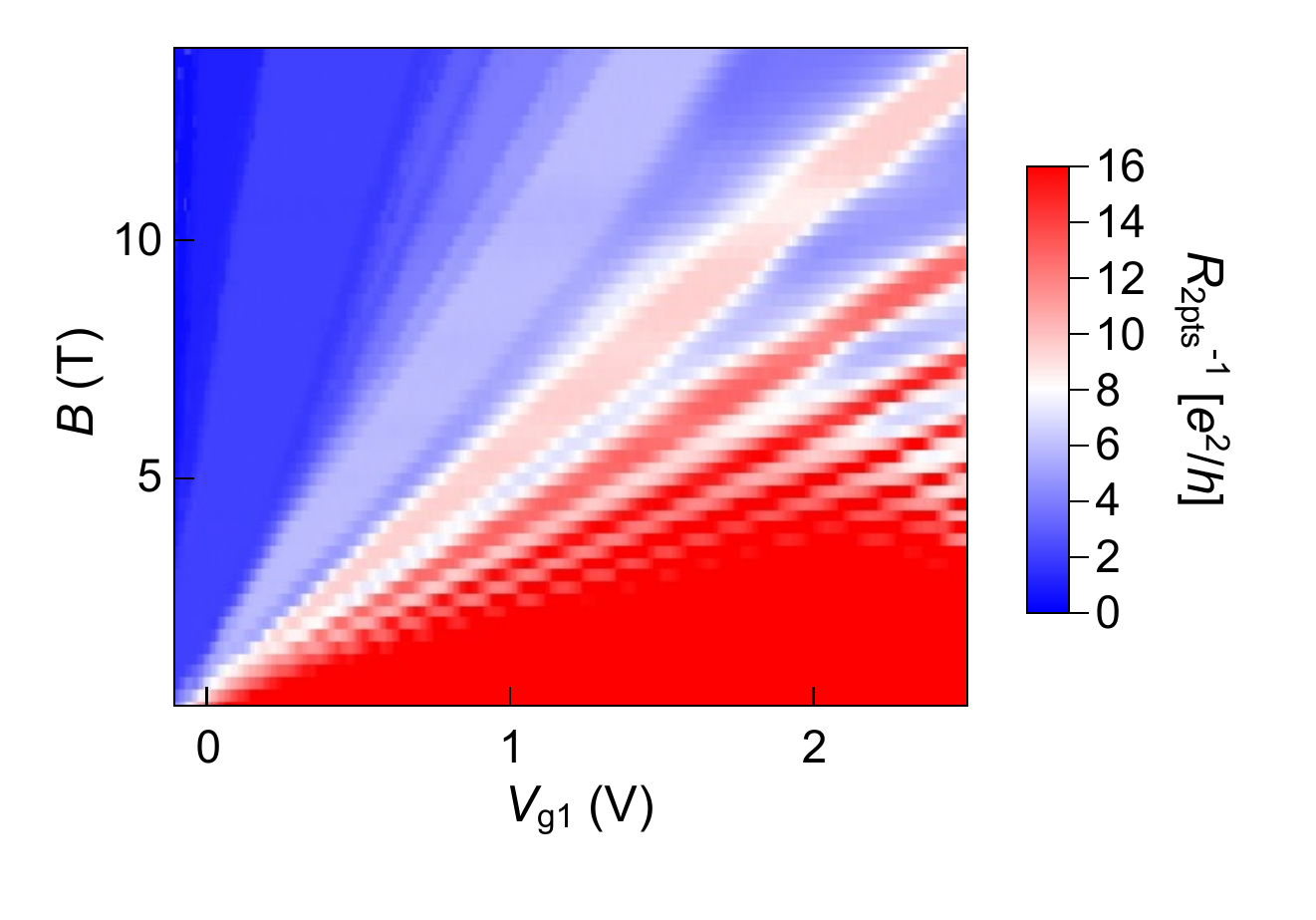}
\caption{\label{figsub-FandiagDeviceA} Landau fan diagram measurement of the 2 point conductance of device A as a function of the magnetic field and graphite back gate voltage $\VgR$, at $T=10~$mK and $\VSi=40~$V, corresponding to the measurement shown in main text Figure 1d.
}
\end{figure}

\begin{figure}[ht!]
\centering
\includegraphics[width=0.98\textwidth]{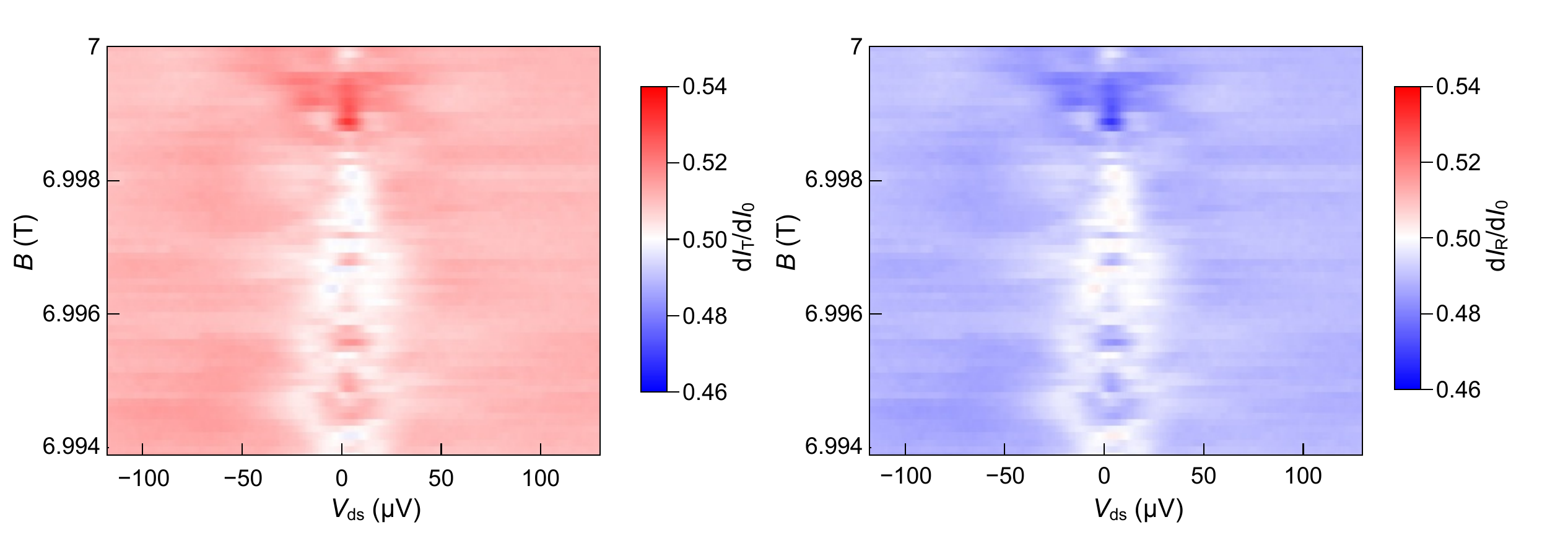}
\caption{\label{figsub-GvsVdsandB-deviceA} Measured transmitted (left) and reflected (right) signals versus drain source voltage and magnetic field in device A, for $\nu_1=\nu_2=2$ and $\VSi=60~$V, corresponding to the data shown in main text Figure 2.
}
\end{figure}

\begin{figure}[ht!]
\centering
\includegraphics[width=0.47\textwidth]{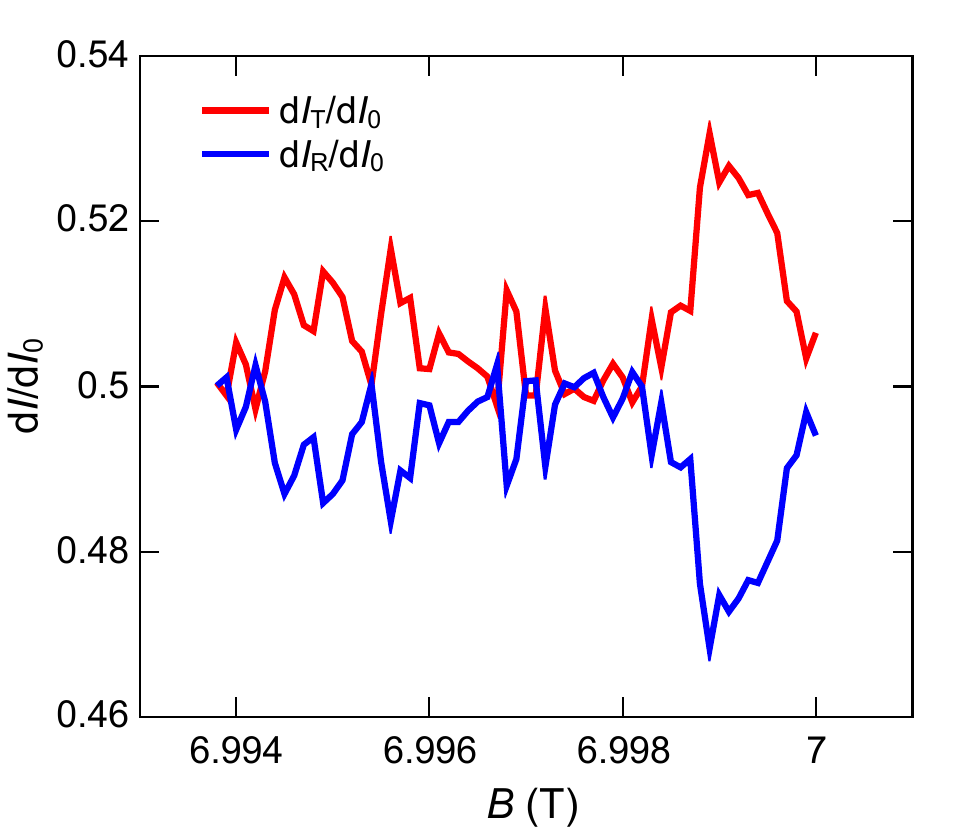}
\caption{\label{figsub-GvsBzerobias-deviceA} Linecut at zero drain-source voltage of the measurements of the reflected (blue) and transmitted (red) signals versus magnetic field in device A, for $\nu_1=\nu_2=2$ and $\VSi=60~$V, corresponding to the data shown in main text Figure 2 and Supplementary Figure~\ref{figsub-GvsVdsandB-deviceA}.
}
\end{figure}

\clearpage

\section{Additional data - device B}


\begin{figure}[ht!]
\centering
\includegraphics[width=0.57\textwidth]{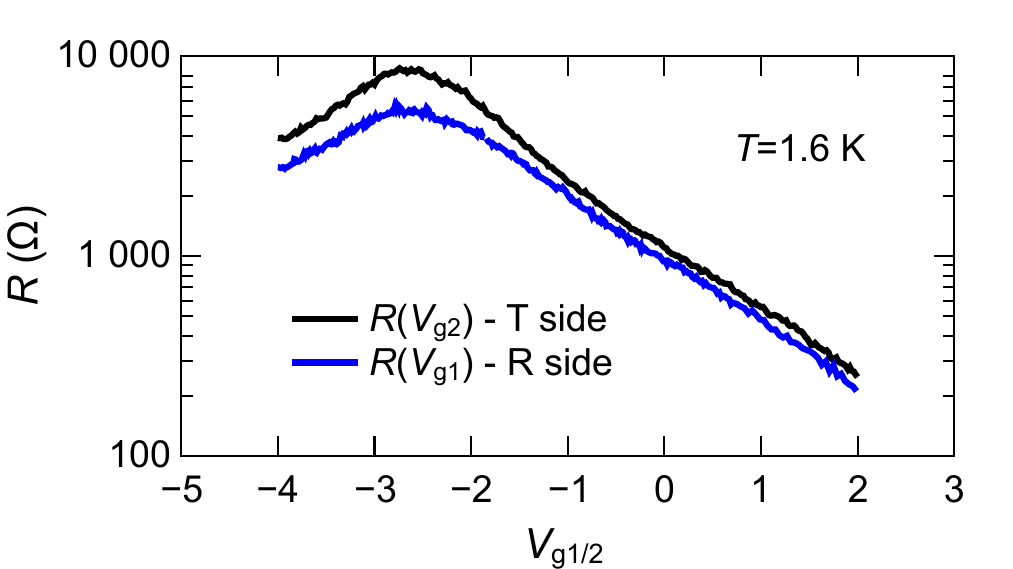}
\caption{\label{figsub-RvsVgDeviceB} Measurement of the two point resistance in device B (first cooldown) versus graphite gate voltages, at $B=0~$T, $T=1.6~$K, and $\VSi=50~$V.
}
\end{figure}

\begin{figure}[ht!]
\centering
\includegraphics[width=0.97\textwidth]{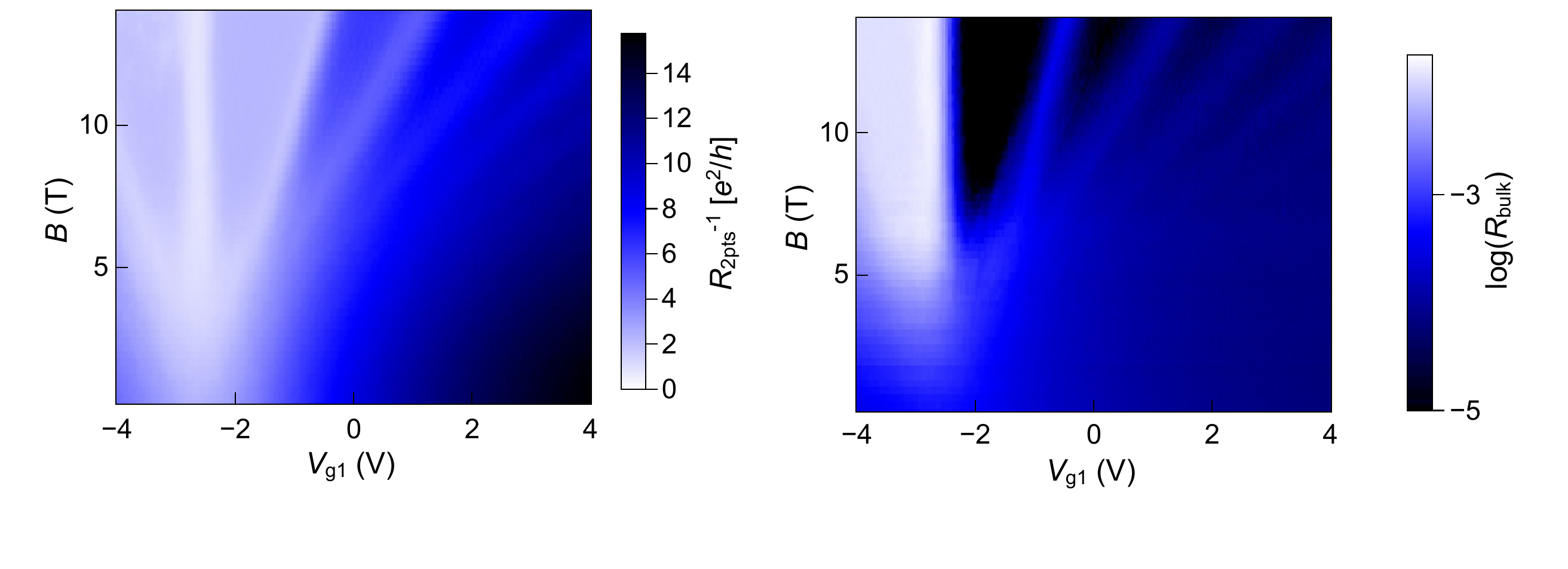}
\caption{\label{figsub-FandiagDeviceB} Landau fan diagram measurement of the 2 point conductance (left) and the bulk resistance (right) of device B (first cooldown) as a function of the magnetic field and graphite back gate voltage $\VgR$, at $T=25~$mK and $\VSi=50~$V.
}
\end{figure}

\begin{figure}[ht]
\centering
\includegraphics[width=0.47\textwidth]{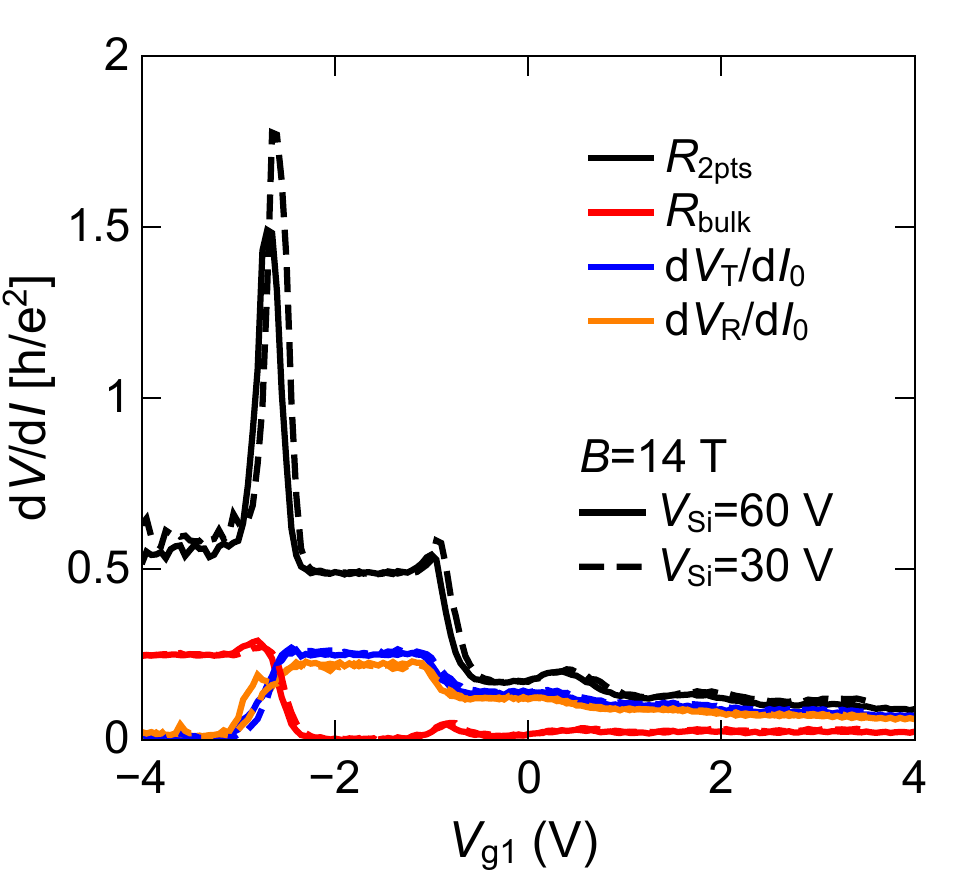}
\caption{\label{figsub-RvsVg14TDeviceB} Measurement of the differential resistances in device B (first cooldown) as a function of the graphite back gate voltage $\VgR$, at $B=14~$T, $T=20~$mK, and $\VgT=-1.9~$V, corresponding to $\nu_2=2$. Black: 2-point resistance $R_\mathrm{2pts}$, red: bulk resistance $R_\mathrm{bulk}$, blue: transmitted transresistance $\dVT/\dIzero$, orange: reflected transresistance $\dVR/\dIzero$. Full lines: $\VSi=60~$V; dashed lines: $\VSi=30~$V.
}
\end{figure}

\begin{figure}[ht]
\centering
\includegraphics[width=0.97\textwidth]{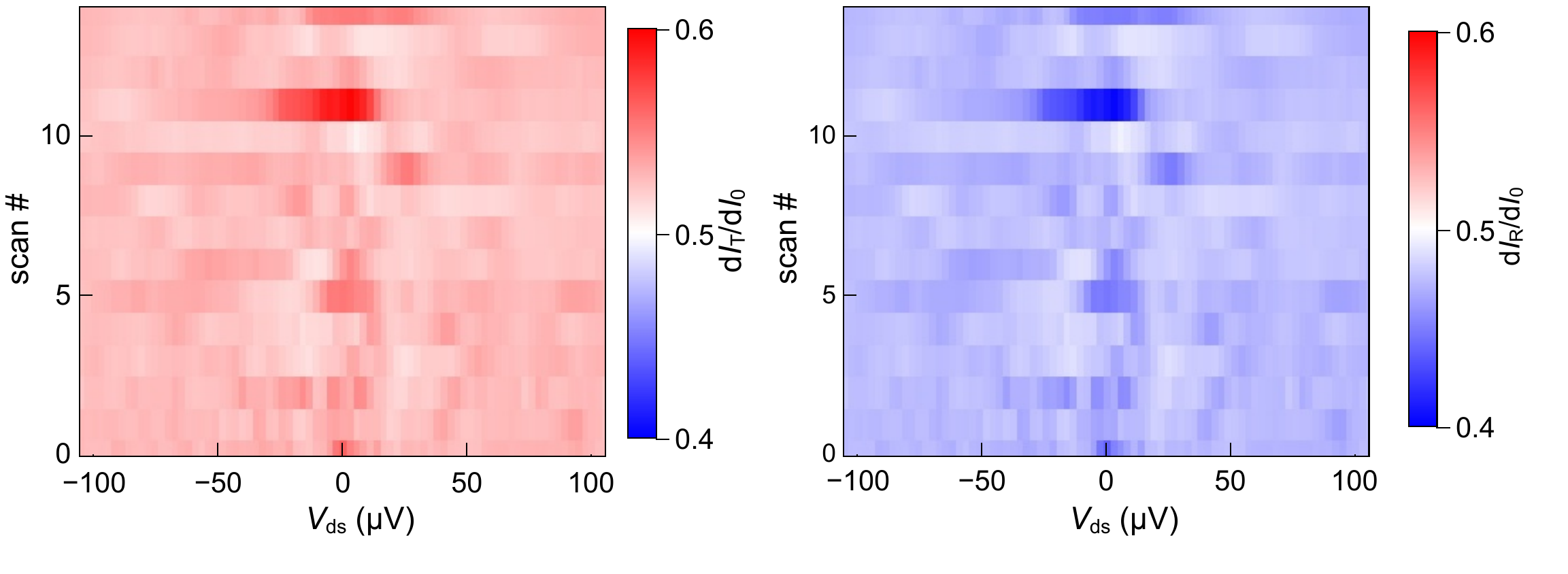}
\caption{\label{figsub-GvsVdsandTimeDeviceB} Measurement of the transmitted (left) and reflected (right) signals in device B (first cooldown) as a function of the drain source voltage, for 15 consecutive scans, at $B=11~$T, $\VSi=60~$V, and $\nu_1=\nu_2=2$. Each scan is about ten minutes long, during which the persistent magnetic field slowly drifts, changing and smearing the interference patterns. The data appearing in main text Figure 4 corresponds to scan number 11 on this plot.
}
\end{figure}
